\documentclass[10p, conference]{IEEEtran}
\usepackage{cite}
\newcommand{\CLASSINPUTtoptextmargin}{2.4cm}
\usepackage{amsmath,amssymb,amsfonts}
\usepackage{graphicx}
\usepackage{textcomp}
\usepackage{xcolor}
\usepackage{soul}
\usepackage{halloweenmath}
\usepackage{comment}
\usepackage{algorithm}
\usepackage{algpseudocode}
\usepackage{comment}
\usepackage{svg}
\usepackage{multirow}
\usepackage{hyperref}
\def\BibTeX{{\rm B\kern-.05em{\sc i\kern-.025em b}\kern-.08em
    T\kern-.1667em\lower.7ex\hbox{E}\kern-.125emX}}
\algrenewcommand\algorithmicindent{1.0em}

\usepackage{fancyhdr,lipsum}

\fancypagestyle{mahmood}{%
   \fancyhf{} 
   
   \fancyhead[C]{978-1-6654-3540-6/22/\$31.00~\copyright~2022 IEEE. You can use this material personally. Reprinting/republishing this material for advertising or promotion, creating new collective works, reselling or redistributing to servers or lists, or using any copyrighted component in other works require IEEE permission. DOI: \href{https://doi.org/10.1109/GLOBECOM48099.2022.10001109}{10.1109/GLOBECOM48099.2022.10001109}}
}%

\makeatletter
\let\ps@IEEEtitlepagestyle\ps@mahmood
\makeatother

\begin{document}

\title{Near-optimal Cloud-Network Integrated Resource Allocation for Latency-Sensitive B5G}

\author{\IEEEauthorblockN{Masoud Shokrnezhad and Tarik Taleb}
\IEEEauthorblockA{Centre for Wireless Communications (CWC)\\ Oulu University,
Oulu, Finland \\
\{masoud.shokrnezhad; tarik.taleb\}@oulu.fi}
}

\maketitle

 


\begin{abstract}
Nowadays, while the demand for capacity continues to expand, the blossoming of Internet of Everything is bringing in a paradigm shift to new perceptions of communication networks, ushering in a plethora of totally unique services. To provide these services, Virtual Network Functions (VNFs) must be established and reachable by end-users, which will generate and consume massive volumes of data that must be processed locally for service responsiveness and scalability. For this to be realized, a solid cloud-network Integrated infrastructure is a necessity, and since cloud and network domains would be diverse in terms of characteristics but limited in terms of capability, communication and computing resources should be jointly controlled to unleash its full potential. Although several innovative methods have been proposed to allocate the resources, most of them either ignored network resources or relaxed the network as a simple graph, which are not applicable to Beyond 5G because of its dynamism and stringent QoS requirements. This paper fills in the gap by studying the joint problem of communication and computing resource allocation, dubbed CCRA, including VNF placement and assignment, traffic prioritization, and path selection considering capacity constraints as well as link and queuing delays, with the goal of minimizing overall cost. We formulate the problem as a non-linear programming model, and propose two approaches, dubbed B\&B-CCRA and WF-CCRA respectively, based on the Branch \& Bound and Water-Filling algorithms. Numerical simulations show that B\&B-CCRA can solve the problem optimally, whereas WF-CCRA can provide near-optimal solutions in significantly less time.
\end{abstract}

\begin{IEEEkeywords}
Beyond 5G, 6G, Computing First Networking, Cloud-Network Integration, Resource Allocation, Path Selection, Traffic Prioritization, VNF Placement, and Optimization Theory.
\end{IEEEkeywords}

\vspace{-0.6cm}
\section{Introduction}\label{S_INTRODUCTION}

As of today, the major reason for the evolution of networks has been a surge in data flow, which has resulted in a continuous 1000x gain in capacity. While this demand for capacity will continue to expand, the blossoming of Internet of Everything is forging a paradigm shift to new-born perceptions bringing a range of entirely novel services with rigorous deterministic criteria, such as connected robotics, smart healthcare, autonomous transportation, and extended reality  \cite{bhat_6g_2021}. The provision of these services will be accomplished by establishing several functional components, namely Virtual Network Functions (VNFs), which will generate and consume vast amounts of data that must be processed locally for service responsiveness and scalability.

A distributed cloud architecture is critical in these situations \cite{corneo_surrounded_2021}, which could be realized through a solid cloud-network integrated infrastructure built of distinct domains in Beyond 5G (B5G) \cite{yang_urllc_2021}. These domains can be distinguished by the technology employed, including radio access, transport, and core networks, as well as edge, access, aggregation, regional, and central clouds. Additionally, these resources can be virtualized through the use of technologies such as Network Function Virtualization (NFV), which allows for the creation of isolated virtual entities atop this physical infrastructure. Since distributed cloud and network domains would be diverse in terms of characteristics but limited in terms of capability, communication and computing resources should be jointly allocated, prioritized, and scheduled to ensure maximum Quality of Service (QoS) satisfaction while maximizing resource sharing and maintaining the system in a deterministic state, resulting in energy savings as one of the most significant examples of cost minimization objectives \cite{li_cognitive_2021}. 

The joint problem of resource allocation in cloud-network integrated infrastructures has been extensively studied in the literature. In \cite{emu_latency_2020}, the authors examined the VNF placement problem as an Integer Linear Programming (ILP) model that assures the minimal End-to-End (E2E) latency while maintaining QoS requirements by not exceeding an acceptable latency violation limit. They suggested an approach based on neural networks and demonstrated that it can produce near-optimal solutions in a timely manner. The authors in \cite{vasilakos_towards_2021} investigated the same problem and proposed a hierarchical reinforcement learning method that includes local level prediction modules as well as a global learning component. They demonstrated that their method significantly outperforms conventional approaches. The similar topic was investigated in \cite{sami_demand-driven_2021}, with the goal of maximizing the number of accepted requests, and a Markov decision process design was presented. They asserted that the proposed method provides efficient placements. 

Although innovative approaches are presented in \cite{emu_latency_2020, vasilakos_towards_2021, sami_demand-driven_2021} to address computing resource constraints, the network is solely viewed as a pipeline in these papers with no cognitive ability to the cloud domains. However, there are also some studies in the literature that have been concentrating on communication and computing resources jointly. In \cite{kuo_deploying_2018}, the joint problem of VNF placement and path selection was investigated to better utilize the network resources, and a heuristic approach was proposed to tackle it. The authors of \cite{mada_latency-aware_2020} and \cite{zhang_adaptive_2019} addressed the problem of VNF placement with the goal of maximizing the sum rate of accepted requests. In \cite{mada_latency-aware_2020}, an optimization solver is used to find the optimal solution, while the solution approach offered in \cite{zhang_adaptive_2019} is a heuristic strategy. The authors of \cite{yuan_toward_2020} formulated the latency-optimal placement of functions as an ILP problem and proposed a genetic meta-heuristic algorithm to solve it. In \cite{gao_cost-efficient_2020}, to reduce the cost of computing resources, the problem of VNF placement and scheduling was addressed, and a latency-aware heuristic algorithm was devised. 

The methods proposed in the cited studies are clearly effective in addressing the resource allocation problem. They cannot, however, be used in B5G systems. Because of the stringent QoS requirements in the delay-reliability-rate space \cite{alwis_survey_2021}, the large number of concurrent services and requests, and the ever-changing dynamics of both infrastructure and end-user service consumption behavior across time and space, every detail of communication and computing resources should be decided and controlled towards achieving a deterministic B5G system \cite{yang_urllc_2021}. In \cite{kuo_deploying_2018}, latency-related constraints and requirements are simply disregarded. Despite the fact that delay is addressed in the rest of these studies, they simplified it to be a link feature, and queuing delay is completely eliminated. Furthermore, path selection is ignored in \cite{mada_latency-aware_2020, zhang_adaptive_2019, yuan_toward_2020}, and cost optimization is overlooked in \cite{zhang_adaptive_2019, yuan_toward_2020}.

This paper fills in the gap in the existing works by studying the joint problem of allocating communication and computing resources, including VNF placement and assignment, traffic prioritization, and path selection while taking into account capacity constraints as well as link and queuing delays, with the goal of minimizing overall cost. Our main contributions in this paper are as follows:
\begin{itemize}
    \item Formulating the joint resource allocation problem of the cloud-network integrated infrastructure as a Mixed Integer Non-Linear Programming (MINLP) problem.
    \item Proposing a method based on Branch \& Bound (B\&B) algorithm to find the optimal solution of the problem.
    \item Devising a heuristic approach based on the Water-Filling (WF) algorithm in order to identify near-optimal solutions to the problem.
\end{itemize}

The reminder of this paper is organized as follows. Section \ref{S_SYSTEMMODEL} introduces the system model. Formulating the resource allocation problem is provided in Section \ref{S_PROBLEMDEFINITION}. Next, the B\&B and heuristic approaches are provided in Sections \ref{S_BBCCRA} and \ref{S_WFCCRA}, respectively. Numerical results are illustrated and analyzed in Section \ref{S_SIMULATIONRESULTS}, followed by concluding remarks in Section \ref{S_CONCLUSION}.

\section{System Model}\label{S_SYSTEMMODEL}
\begin{figure}[t!]\centering
\includegraphics[width=3in]{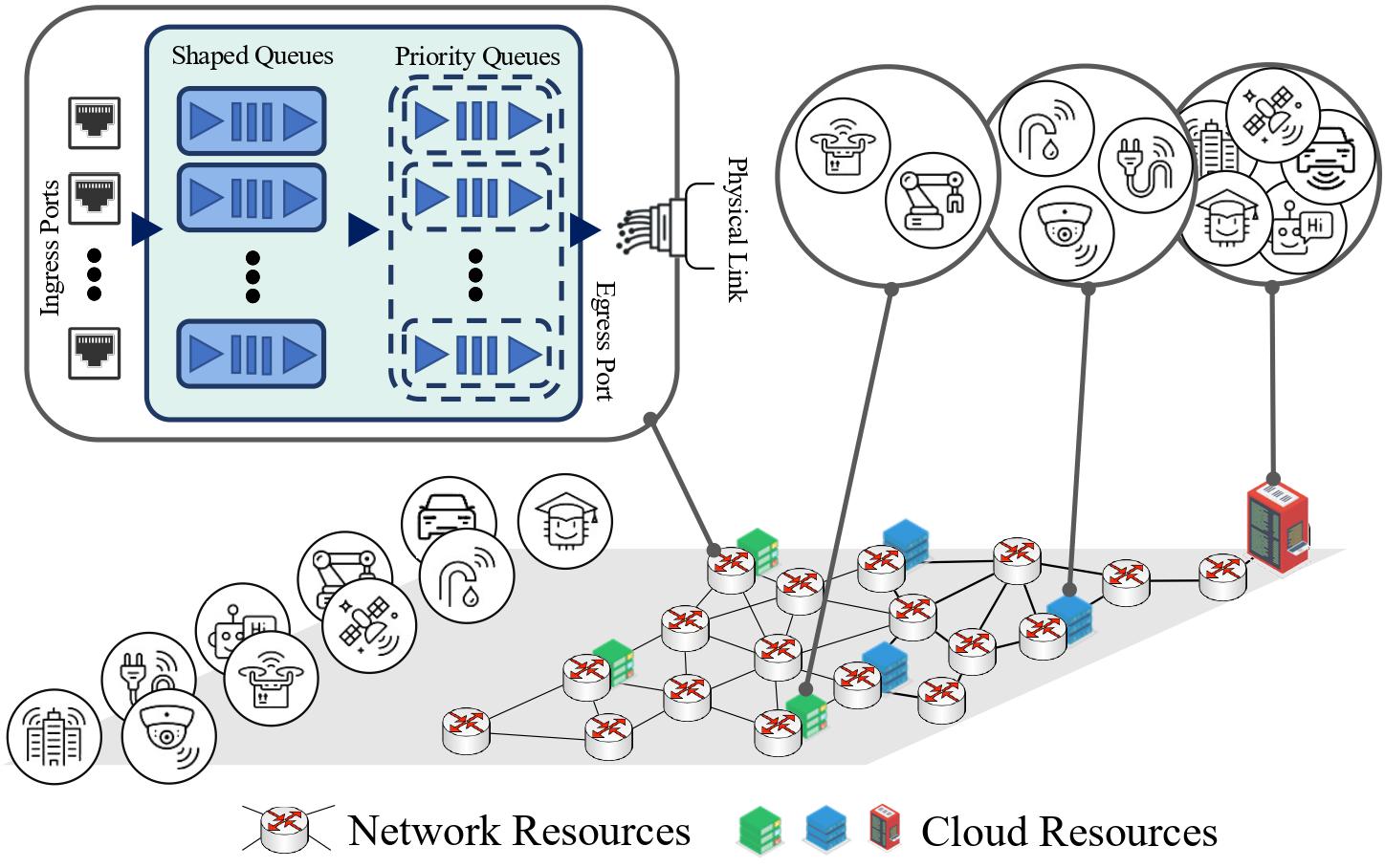}
  \caption{System model}
  \label{fig_01}
\end{figure}

In the following, we describe the main components of the system studied in this paper: infrastructure, services, and requests. The system model is also depicted in Fig \ref{fig_01}.

\subsection{Infrastructure Model}\label{SS_NETWORKMODEL}
The considered infrastructure is composed of the access and core network domains (non-radio domains) consisting of $\mathcal{V}$ nodes, $\mathcal{L}$ links, and $\mathcal{P}$ paths denoted by $\mathcal{G} = \langle \boldsymbol{\mathcal{V}},\boldsymbol{\mathcal{L}}, \boldsymbol{\mathcal{P}} \rangle$. $\boldsymbol{\mathcal{V}} = \{1,2,...,\mathcal{V}\}$ is the set of nodes. $\boldsymbol{\mathcal{L}} \subset \{l:(v,v') | v,v' \in \boldsymbol{\mathcal{V}}\}$ indicates the set of links, and for each $l$, its bandwidth capacity is constrained by $\widehat{B_l}$, and it costs $\Xi_l$ per capacity unit. $\boldsymbol{\mathcal{P}} = \{p:(\vdash_p, \dashv_p) | p \subset \boldsymbol{\mathcal{L}}\}$ denotes the set of all paths in the network, where $\vdash_p$ and $\dashv_p$ are the head and tail nodes of path $p$, and $l'_{l,p}$ is a binary constant equal to $1$ if path $p$ contains link $l$. 

Each node in the network is an IEEE 802.1 Time-Sensitive Networking (TSN) device comprising an IEEE 802.1 Qcr Asynchronous Traffic Shaper (ATS) at each of their egress ports. An ATS consists of two hierarchical queuing steps \cite{specht_urgency-based_2016}: interleaved shaping, and scheduling through a set of prioritized queues. We consider $\boldsymbol{\mathcal{K}} = \{1,2,...,\mathcal{K}\}$ as the set of priority levels and assume that $k_r$ is the assigned priority of the traffic associated with request $r$, and the size of the shaping queues for priority level $k$ is the same and equal to $\widehat{\mathcal{T}_k}$. Note that lower levels have higher priorities. Moreover, each node $v$ is equipped with computing resources as one of the prospective hosts to deploy service VNFs and limited to a predefined capacity threshold $\widehat{\zeta_v}$ which costs $\Psi_v$ per capacity unit. 

It is worth mentioning that the network is divided into a number of tiers, with nodes distributed across them so that the entry nodes of requests are located in tier $0$. The higher the tier index, the greater the capacity of the associated nodes, and the lower their cost. In other words, the nodes closest to end-users (or to the nodes that serve as entry points) are provisioned with high-cost, limited-capacity computing facilities, while low-cost, high-capacity depots are deployed in the core.

\subsection{Service Model}\label{SS_SERVICEMODEL}
The set of services obtainable to order is dubbed by $\boldsymbol{\mathcal{S}}=\{1,2,...,\mathcal{S}\}$, where $\mathcal{S}$ indicates the number of services. If an end-user requests a service, its VNF has to be replicated in the network-embedded computing resources. Each VNF is empowered to serve more than one request, and $\widehat{\mathcal{C}_s}$ indicates the maximum capacity of each VNF of service $s$.

\subsection{Request Model}\label{SS_REQUESTMODEL}
The set of requests asking for services is represented by $\boldsymbol{\mathcal{R}}=\{1,2,...,\mathcal{R}\}$, where $\mathcal{R}$ is the number of requests. Each request $r$ arrives in the network through node $v_r$, one of the nodes through which the infrastructure connects to the radio access network, and intends a service $s_r$ specifying its minimum necessitated service capacity, network bandwidth, and maximum tolerable delay, indicated by $\widetilde{\mathcal{C}_r}$, $\widetilde{\mathcal{B}_r}$, and $\widetilde{\mathcal{D}_r}$, respectively. In addition, $\widetilde{\mathcal{T}_r}$ and $\widetilde{\mathcal{H}_r}$, denoting the burstiness of traffic and the largest packet size for request $r$, are also assumed to be known a priori. Utilizing historical data along with predictive data analytics methods is one of the viable options for obtaining such accurate and realistic statistical estimates of traffic.

\section{Problem Definition}\label{S_PROBLEMDEFINITION}

In this section, the joint problem of VNF placement and assignment, traffic prioritization, and path selection is described. The constraints and objective function are formulated as a MINLP problem in what follows, and the problem is stated at the end of the section.

\subsection{VNF Placement and Assignment Constraints}\label{SS_SERVICEPLACEMENTANDASSIGNMENTCONSTRAINTS}
To begin, each request must be assigned a single node as its service location (C1). This assignment is acceptable if the assigned node hosts a VNF for the requested service (C2). Following that, it must be ensured that the assigned requests do not violate the capacity constraints of VNFs and nodes (C3 and C4). Note that the capacity constraints are intrinsically linked to avoiding congestion and ensuring the system's reliability. The formulation is as follows:
\begin{align}\label{CS_SERVICEPLACEMENTANDASSIGNMENTCONSTRAINTS}
    &\footnotesize \sum\nolimits_{\boldsymbol{\mathcal{V}}} g_{r,v} = 1, \forall r \in  \boldsymbol{\mathcal{R}},\tag{\footnotesize C1} 
    \\
    &\footnotesize g_{r,v} \leq z_{s_r,v}, \forall r \in  \boldsymbol{\mathcal{R}}, \forall v \in  \boldsymbol{\mathcal{V}},\tag{\footnotesize C2} 
    \\
    &\footnotesize \sum\nolimits_{\{r | r \in  \boldsymbol{\mathcal{R}} \wedge s_r = s\}} \widetilde{\mathcal{C}_r} g_{r,v} \leq \widehat{\mathcal{C}_s}, \forall v \in  \boldsymbol{\mathcal{V}}, \forall s \in \boldsymbol{\mathcal{S}},\tag{\footnotesize C3}  
    \\
    &\footnotesize  \sum\nolimits_{\boldsymbol{\mathcal{S}}} \widehat{\mathcal{C}_s} z_{s,v} \leq \widehat{\zeta_v}, \forall v \in \boldsymbol{\mathcal{V}},\tag{\footnotesize C4}
\end{align}
where $g_{r,v}$ and $z_{s,v}$ are binary variables. $g_{r,v}$ is $1$ if node $v$ is selected as the service node of request $r$, and $z_{s,v}$ is $1$ if service $s$ is replicated on node $v$. 

\subsection{Traffic Prioritization and Path Selection Constraints}\label{SS_PRIORITYASSIGNMENTANDPATHSELECTIONCONSTRAINTS}
First, we must ensure that each request is assigned to exactly one priority level (C5). Then, the request and reply paths of each request are determined (C6 and C7). For each request, a single request path is chosen that starts at the request's entry node and ends at the request's VNF node. The reply path follows the same logic but in reverse order. The following two constraints guarantee that the two paths are chosen on the priority level assigned to each request (C8 and C9). Finally, the constraints maintaining the maximum capacity of links and shaping queues are enforced (C10 and C11). With C10, the sum of the required bandwidth for all requests whose request or reply path, or both, contains link $l$ is guaranteed to be less than or equal to the link's capacity, and in C11, the capacity of shaping queues is guaranteed in the same way for each link and each priority level. The set includes:
\begin{align}\label{CS_PRIORITYASSIGNMENTANDPATHSELECTIONCONSTRAINTS1}
    &\footnotesize \sum\nolimits_{\boldsymbol{\mathcal{K}}} \varrho_{r,k} = 1, \forall r \in  \boldsymbol{\mathcal{R}},\tag{\footnotesize C5} 
    \\
    &\footnotesize \sum\nolimits_{\{p | p \in \boldsymbol{\mathcal{P}} \wedge \vdash_p = v_r \wedge \dashv_p = v \}, \boldsymbol{\mathcal{K}}} \overrightarrow{f_{r,p,k}} = g_{r,v}, \forall r \in  \boldsymbol{\mathcal{R}}, \forall v \in \boldsymbol{\mathcal{V}},\tag{\footnotesize C6} 
    \\
    &\footnotesize \sum\nolimits_{\{p | p \in \boldsymbol{\mathcal{P}} \wedge \vdash_p = v \wedge \dashv_p = v_r \}, \boldsymbol{\mathcal{K}}} \overleftarrow{f_{r,p,k}} = g_{r,v}, \forall r \in  \boldsymbol{\mathcal{R}}, \forall v \in \boldsymbol{\mathcal{V}},\tag{\footnotesize C7}
    \\
    &\footnotesize \sum\nolimits_{\boldsymbol{\mathcal{P}}} \overrightarrow{f_{r,p,k}} = \varrho_{r,k}, \forall r \in  \boldsymbol{\mathcal{R}}, \forall k \in \boldsymbol{\mathcal{K}},\tag{\footnotesize C8}
    \\
    &\footnotesize \sum\nolimits_{\boldsymbol{\mathcal{P}}} \overleftarrow{f_{r,p,k}} = \varrho_{r,k}, \forall r \in  \boldsymbol{\mathcal{R}}, \forall k \in \boldsymbol{\mathcal{K}},\tag{\footnotesize C9}
    \\
    &\footnotesize \sum\nolimits_{\boldsymbol{\mathcal{R}}} \widetilde{\mathcal{B}_r} \sum\nolimits_{\boldsymbol{\mathcal{P}}, \boldsymbol{\mathcal{K}}} l'_{l,p}(\overrightarrow{f_{r,p,k}} + \overleftarrow{f_{r,p,k}}) \leq \widehat{B_{l}}, \forall l \in  \boldsymbol{\mathcal{L}}, \tag{\footnotesize C10} 
    \\
    &\footnotesize \sum\nolimits_{\boldsymbol{\mathcal{R}}} \widetilde{\mathcal{T}_r} \sum\nolimits_{\boldsymbol{\mathcal{P}}} l'_{l,p}(\overrightarrow{f_{r,p,k}} + \overleftarrow{f_{r,p,k}}) \leq \widehat{\mathcal{T}_k}, \forall k \in \boldsymbol{\mathcal{K}}, \forall l \in \boldsymbol{\mathcal{L}},\tag{\footnotesize C11}
\end{align}
where $\varrho_{r,k}$ is a binary variable representing the assigned priority level of request $r$, and $\overrightarrow{f_{r,p,k}}$ and $\overleftarrow{f_{r,p,k}}$ are binary variables that reflect the request and reply paths for request $r$ on priority level $k$, respectively.

\subsection{Delay Constraints}\label{SS_DELAYCONSTRAINTS}
The final set guarantees the minimum delay requirement of requests as follows:
\begin{align}\label{CS_DELAYCONSTRAINTS_P1P2}
    &\footnotesize D_{r, s_r} = \frac{\widetilde{\mathcal{H}_r}}{\widetilde{\mathcal{C}_r}}, \forall r \in \boldsymbol{\mathcal{R}}, \tag{\footnotesize C12}\\
    & \footnotesize D_{r,k,l} = \frac{\sum\nolimits_{\boldsymbol{\mathcal{R}_1}} \widetilde{\mathcal{T}_{r'}} + \bigwedge\nolimits_{\boldsymbol{\mathcal{R}_2}} \widetilde{\mathcal{H}_{r'}}}{\widehat{B_{l}} - \sum\nolimits_{\boldsymbol{\mathcal{R}3}} \widetilde{\mathcal{B}_{r'}}} + \frac{\widetilde{\mathcal{H}_{r}}}{\widehat{B_{l}}}, \;\forall k \in \boldsymbol{\mathcal{K}}, \forall l \in \boldsymbol{\mathcal{L}}, \tag{\footnotesize C13} \\
    &\footnotesize \sum\nolimits_{\boldsymbol{\mathcal{P}}, \boldsymbol{\mathcal{L}}, \boldsymbol{\mathcal{K}}} D_{r,k,l} l'_{l,p}(\overrightarrow{f_{r,p,k}} + \overleftarrow{f_{r,p,k}}) + D_{r, s_r}, \forall r \in \boldsymbol{\mathcal{R}}, \tag{\footnotesize C14} \\
    &\footnotesize D_{r} \leq \widetilde{\mathcal{D}_r}, \forall r \in \boldsymbol{\mathcal{R}},\tag{\footnotesize C15}
\end{align}
where $D_{r,k,l}$, $D_{r, s_r}$ and $D_{r}$ are continues variables denoting the delay experienced by a given ﬂow of request $r$ associated to priority level $k$ passing through ATS-based link $l$ \cite{specht_urgency-based_2016}, its computing delay, and the corresponding E2E delay calculated as the sum of the delays on the links that comprise both paths of the request and its computing delay. Besides, $\bigwedge$ is a function which returns the max value over the given set, $\boldsymbol{\mathcal{R}_1}$ represents $\{r' | r' \in \boldsymbol{\mathcal{R}} \wedge k_{r'} \leq k \wedge l'_{l,p}(\overrightarrow{f_{r',p, k_{r'}}} + \overleftarrow{f_{r',p, k_{r'}}}) > 0\}$, $\boldsymbol{\mathcal{R}_2}$ is $\{r' | r' \in \boldsymbol{\mathcal{R}} \wedge k_{r'} > k \wedge l'_{l,p}(\overrightarrow{f_{r',p,k_{r'}}} + \overleftarrow{f_{r',p,k_{r'}}}) > 0 \}$, and $\boldsymbol{\mathcal{R}_3}$ denotes $\{r' | r' \in \boldsymbol{\mathcal{R}} \wedge k_{r'} < k \wedge l'_{l,p}(\overrightarrow{f_{r',p,k_{r'}}} + \overleftarrow{f_{r',p,k_{r'}}}) > 0\}$. In other words, these sets represent requests that share the same link as $r$, whereas $\boldsymbol{\mathcal{R} 1}$ includes requests with a higher or equal priority, $\boldsymbol{\mathcal{R}_2}$ contains requests with a lower priority, and $\boldsymbol{\mathcal{R}_3}$ shares requests with a higher priority.

\subsection{Objective Function}\label{SS_OF}
The objective function is to minimize the total cost of allocated resources, that is:
\begin{equation*}\label{OBJECTIVEFUNCTION}
    \footnotesize \sum\nolimits_{\boldsymbol{\mathcal{R}}, \boldsymbol{\mathcal{V}}} \Psi_v g_{r,v} + \sum\nolimits_{\boldsymbol{\mathcal{R}}, \boldsymbol{\mathcal{L}}} \Xi_{l} \sum\nolimits_{\boldsymbol{\mathcal{P}}, \boldsymbol{\mathcal{K}}} l'_{l,p}(\overrightarrow{f_{r,p,k}} + \overleftarrow{f_{r,p,k}}). \tag{OF}
\end{equation*}
Given that the cost of links and computing resources would be easily linked to their energy consumption, the objective function could be interpreted as minimizing energy consumption as a critical goal of developing B5G sustainable communication and computing systems \cite{bhat_6g_2021}.

\subsection{Problem}\label{SS_PROBLEM}
Considering the constraints and objective function, the problem of Communication and Computing Resource Allocation (CCRA) is:
\begin{equation}\label{EQ_CCRA}
    \footnotesize \text{CCRA: } \textit{ min } \text{OF} \textit{ s.t. } \text{C1 - C15.} 
\end{equation}

\section{B\&B-CCRA}\label{S_BBCCRA}

\begin{algorithm}[b!]
\caption{B\&B-CCRA}\label{ALG_BB}
\begin{algorithmic}[1]
\State $\boldsymbol{\mathcal{N}} \gets \{N_1\}$, $\eta^\star \gets +\infty$, $t \gets 0$
\State \textbf{while} $\boldsymbol{\mathcal{N}}$ is not empty \textbf{do}
\State \textbar \textbf{} $t \gets t+1$
\State \textbar \textbf{} $N_t \gets$ selects a node form $\boldsymbol{\mathcal{N}}$
\State \textbar \textbf{} $\boldsymbol{\mathcal{N}} \gets \boldsymbol{\mathcal{N}}$\textbackslash$\{N_t\}$
\State \textbar \textbf{} $(\boldsymbol{\mu}^{\star}_t,\boldsymbol{\lambda}^{\star}_t), \phi^{\star}_t \gets$ solve the relaxed problem of $\Phi_t$
\State \textbar \textbf{ if} $\boldsymbol{\mu}^{\star}_t$ is integer for all elements \textbf{then}
\State \textbar \textbf{} \textbar \textbf{} $\eta^\star \gets min(\eta^\star,\phi^\star_t)$
\State \textbar \textbf{ else if} $\phi^{\star}_t < \eta^\star$ is preserved \textbf{then}
\State \textbar \textbf{} \textbar \textbf{} $N^1_t,N^2_t \gets$ two children of $N_t$
\State \textbar \textbf{} \textbar \textbf{} $\boldsymbol{\mathcal{N}} \gets \boldsymbol{\mathcal{N}} \cup \{N^1_t,N^2_t\}$
\end{algorithmic}
\end{algorithm}

The problem specified in (\ref{EQ_CCRA}) is NP-hard (the multidimensional knapsack problem \cite{kellerer_multidimensional_2004} can be reduced to it, as detailed in \cite{faticanti_cutting_2018}), and finding its optimal solution in polynomial time is mathematically intractable. One potential strategy for addressing such a problem is to restrict its solution space using the B\&B algorithm, which relaxes and solves the problem to obtain lower bounds, and then improves the bounds using mathematical techniques to reach acceptable solutions. The method is described in Algorithm \ref{ALG_BB}. In this algorithm, the solution space is discovered by maintaining an unexplored candidate list $\boldsymbol{\mathcal{N}}=\{N_t | t \geq 1\}$, where each node $N_t$ contains a problem, denoted by $\Phi_t$, and $t$ is the iteration number. This list only contains the root candidate $N_1$ at the beginning with the primary problem to be solved. To reduce its enormous computational complexity, instead of directly applying the B\&B algorithm to CCRA, we consider its integer linear transformation as the problem of $N_1$.

CCRA comprises non-linear constraints C13 and C14. To linearize C13, the summations and max function with variable boundaries should be converted to a linear form. A simple, effective technique is to replace each term with an approximated upper bound. Since the aggregated traffic burstiness is bounded by $\widehat{\mathcal{T}_{k}}$ for each priority level $k$ in C11, $\sum_{\boldsymbol{\mathcal{R}_1}} \widetilde{\mathcal{T}_{r'}}$ can be replaced by the sum of this bound for all priority levels greater than or equal to $k$, that is $\sum_{\{k'|k' \leq k\}} \widehat{\mathcal{T}_{k'}}$. In a similar way, we define a new constraint (C13$'$) for the aggregated bandwidth allowed on priority level $k$ over link $l$, dubbed $\widehat{f_{l,k}}$, and replace the sum of allocated bandwidths with $\sum_{\{k'|k' < k\}} \widehat{f_{l,k'}}$. Besides, the maximum packet size for a particular subset of requests can be replaced by the maximum permitted packet size in the network, denoted by $\widehat{\mathcal{H}}$. Therefore, the followings define the linear transformation of C13:
\begin{align}\label{CS_C11}
    &\footnotesize
    \begin{aligned}
        \sum\nolimits_{\boldsymbol{\mathcal{R}}} \widetilde{\mathcal{B}_r} \sum\nolimits_{\boldsymbol{\mathcal{P}}} l'_{l,p}(\overrightarrow{f_{r,p,k}} + \overleftarrow{f_{r,p,k}}) \leq & \widehat{f_{l,k}}, \forall k \in \boldsymbol{\mathcal{K}}, \forall l \in \boldsymbol{\mathcal{L}},
    \end{aligned}\tag{\footnotesize C13$'$}\\
    &\footnotesize
    \begin{aligned}
        \widehat{D_{r,k,l}} = & \frac{\sum\nolimits_{\{k'|k' \leq k\}} \widehat{\mathcal{T}_{k'}} + \widehat{\mathcal{H}}}{\widehat{B_{l}} - \sum_{\{k'|k' < k\}} \widehat{f_{l,k'}}} + \frac{\widetilde{\mathcal{H}_{r}}}{\widehat{B_{l}}}, \forall r \in \boldsymbol{\mathcal{R}}, \forall k \in \boldsymbol{\mathcal{K}},\\
        &\forall l \in \boldsymbol{\mathcal{L}},
    \end{aligned}\tag{\footnotesize C13$''$} 
\end{align}
where $\widehat{D_{r,k,l}}$ is the delay upper bound for request $r$ on link $l$ with priority level $k$. Since $D_{r, s_r}$ is linear, C14 can be linearized by substituting the actual delay for the upper bound derived in C13$''$, and the new constraint for E2E delay is:
\begin{align}\label{CS_C14}
    &\footnotesize
    \begin{aligned}
        D_{r} = \sum\nolimits_{\boldsymbol{\mathcal{P}},\boldsymbol{\mathcal{L}},\boldsymbol{\mathcal{K}}} \widehat{D_{r,k,l}} l'_{l,p}(\overrightarrow{f_{r,p,k}} + \overleftarrow{f_{r,p,k}}) + D_{r, s_r}, \forall r \in \boldsymbol{\mathcal{R}}.
    \end{aligned}\tag{\footnotesize C14$'$} 
\end{align}
Given this, the linear transformation of CCRA, dubbed LiCCRA, is as follows:
\begin{align*}\label{EQ_LiCCRA}
\footnotesize \text{LiCCRA:} \textit{ min } \text{OF} \textit{ s.t. } &\text{C1 - C12, C13$'$, C13$''$, C14$'$, C15}.\tag{2} 
\end{align*}

Now, with LiCCRA as $\Phi_1$, each iteration of the B\&B algorithm begins with the selection and removal of a candidate from the unexplored list. Then, the problem of this candidate is naturally relaxed and solved, i.e., all the integer variables ($\in \{0,1\}$) are replaced with their continues equivalents restricted by the box constraint ($\in [0,1]$), and the relaxed problem is solved using a Linear Programming (LP) solver to obtain the solution of the relaxed problem $(\boldsymbol{\mu}^{\star}_t,\boldsymbol{\lambda}^{\star}_t)$ and the optimal objective value $\phi^{\star}_t$, where $\boldsymbol{\mu}$ is the relaxed integer variables set, and $\boldsymbol{\lambda}$ is the set of continuous variables. Next, if all relaxed variables have integer values, the obtained objective in this iteration is considered to update the best explored integer solution. Otherwise, a variable index $j$ is selected such that $\boldsymbol{\mu}^{\star}_t[j]$ is fractional, and the feasible constraints set $\pi_t$ is divided into two parts as $\pi_t^1 = \pi_t \cap \{\boldsymbol{\mu}_t[j] \leq \big\lfloor \boldsymbol{\mu}^{\star}_t[j] \big\rfloor\}$ and $\pi_t^2 = \pi_t \cap \{\boldsymbol{\mu}_t[j] \geq \big\lceil \boldsymbol{\mu}^{\star}_t[j] \big\rceil\}$. Then, two problems are formed as $\Phi_t^1 = min \text{ OF } \textit{s.t. } \pi_t^1$ and $\Phi_t^2 = min \text{ OF } \textit{s.t. } \pi_t^2$. Now, two child nodes $N_t^1$ and $N_t^2$, whose problems are $\Phi_t^1$ and $\Phi_t^2$ respectively, are put into the unexplored list. The B\&B algorithm is iterated until $\boldsymbol{\mathcal{N}}$ is empty. 

Alternatively, we can run this algorithm until a desired solving time is reached or an acceptable objective value is acquired. The prime advantage of this algorithm is that it produces at least a lower bound even when the solving time is limited. As a result, it may be used to establish baselines allowing for the evaluation of alternative approaches.

\section{WF-CCRA}\label{S_WFCCRA}

\begin{algorithm}[b!]
\caption{WF-CCRA}\label{ALG_WF}
\vspace{0.1cm}
\begin{algorithmic}[1]
\State initialize variable and parameter vectors
\State $\boldsymbol{\mathcal{R'}} \gets \{\}$, $\boldsymbol{\Omega} \gets \{\}$
\State sort $\boldsymbol{\mathcal{R}}$ in ascending order according to $\widetilde{\mathcal{D}_r}$
\State \textbf{while} $\boldsymbol{\mathcal{R}}$ is not empty \textbf{do}
\State \textbar \textbf{ for} $v \in \boldsymbol{\mathcal{V}}$ \textbf{do}
\State \textbar \textbf{} \textbar \textbf{ if} $z_{s_r,v}==1$ \textbf{and} $\widetilde{\mathcal{C}_r} \leq \widehat{\mathcal{C}_{s_r}}$ on $v$ \textbf{then}
\State \textbar \textbf{} \textbar \textbf{} \textbar \textbf{} $g_{r,v}=1$
\State \textbar \textbf{} \textbar \textbf{ if} $z_{s_r,v} \neq 1$ \textbf{and} $\widehat{\mathcal{C}_{s_r}} \leq \widehat{\zeta_v}$ \textbf{then}
\State \textbar \textbf{} \textbar \textbf{} \textbar \textbf{} $z_{s_r,v}=1$, $g_{r,v}=1$
\State \textbar \textbf{} \textbar \textbf{ else go to the next iteration}
\State \textbar \textbf{} \textbar \textbf{ for} $k \in \boldsymbol{\mathcal{K}}$ \textbf{do}
\State \textbar \textbf{} \textbar \textbf{} \textbar \textbf{} $\varrho_{r,k}=1$
\State \textbar \textbf{} \textbar \textbf{} \textbar \textbf{ for} $p \in \boldsymbol{\mathcal{P}} \wedge \vdash_p = v_r \wedge \dashv_p=v$ \textbf{do}
\State \textbar \textbf{} \textbar \textbf{} \textbar \textbf{} \textbar \textbf{ if} $\widetilde{\mathcal{B}_r} \leq \widehat{\mathcal{B}_{l}} \; \& \; \widetilde{\mathcal{T}_r} \leq \widehat{\mathcal{T}_{k}}$ on $l \; \forall l \in \boldsymbol{\mathcal{L}} \wedge l'_{l,p}=1$ \textbf{then} 
\State \textbar \textbf{} \textbar \textbf{} \textbar \textbf{} \textbar \textbf{} \textbar \textbf{} \tiny $\overrightarrow{f_{r,p,k}} = 1$ \normalsize
\State \textbar \textbf{} \textbar \textbf{} \textbar \textbf{} \textbar \textbf{} \textbar \textbf{ for} $p' \in \boldsymbol{\mathcal{P}} \wedge \vdash_{p'} = v \wedge \dashv_{p'}=v_r $ \textbf{do}
\State  \textbar \textbf{} \textbar \textbf{} \textbar \textbf{} \textbar \textbf{} \textbar \textbf{} \textbar \small \textbf{ if} $\widetilde{\mathcal{B}_r} \leq \widehat{\mathcal{B}_{l}} \; \& \; \widetilde{\mathcal{T}_r} \leq \widehat{\mathcal{T}_{k}}$ on $l \; \forall l \in \boldsymbol{\mathcal{L}} \wedge l'_{l,p'}=1$ \textbf{then} \normalsize
\State \textbar \textbf{} \textbar \textbf{} \textbar \textbf{} \textbar \textbf{} \textbar \textbf{} \textbar \textbf{} \textbar \textbf{} \tiny $\overleftarrow{f_{r,p',k}} = 1$ \normalsize
\State \textbar \textbf{} \textbar \textbf{} \textbar \textbf{} \textbar \textbf{} \textbar \textbf{} \textbar \textbf{} \textbar \textbf{} calculate $D_r$ based on (C14) (or C14$'$)
\State \textbar \textbf{} \textbar \textbf{} \textbar \textbf{} \textbar \textbf{} \textbar \textbf{} \textbar \textbf{} \textbar \textbf{ if} $D_{r} \leq \widetilde{\mathcal{D}_r}$ \textbf{then} 
\State \textbar \textbf{} \textbar \textbf{} \textbar \textbf{} \textbar \textbf{} \textbar \textbf{} \textbar \textbf{} \textbar \textbf{} \textbar \textbf{} $\boldsymbol{\Omega} \gets \boldsymbol{\Omega} \cup \{(z_{s_r,v}, g_{r,v}, \varrho_{r,k}, \overrightarrow{f_{r,p,k}}, \overleftarrow{f_{r,p',k}}) \}$ 
\State \textbar \textbf{} fix assignments of $\textit{argmin}_{\boldsymbol{\Omega}}$ (OF) for $r$
\State \textbar \textbf{} \small update capacities, \normalsize $\boldsymbol{\Omega} \gets \{\}$, $\boldsymbol{\mathcal{R}} \gets \boldsymbol{\mathcal{R}}/\{r\}$, $\boldsymbol{\mathcal{R'}} \gets \boldsymbol{\mathcal{R'}} \cup \{r\}$
\end{algorithmic}
\end{algorithm}

Since the B\&B method searches the problem's solution space for the optimal solution, its complexity can grow up to the size of the solution space in the worst case \cite{pataki_basis_2010}. Given that the size of the solution space in CCRA (or LiCCRA) for each request is $\mathcal{V}^2|\boldsymbol{\mathcal{P}}|^2
\mathcal{K}^3$ considering its integer variables, the problem's overall size is $\mathcal{R!}{\mathcal{V}^2|\boldsymbol{\mathcal{P}}|^2
\mathcal{K}^3}$. Therefore, finding its optimal solution for large-scale instances using B\&B is impractical in a timely manner, and the goal of this section is to devise an efficient approach based on the WF concept in order to identify near-optimal solutions for this problem. 

The WF-CCRA method is elaborated in Algorithm \ref{ALG_WF}. The first step is to initialize the vectors of parameters and variables used in (\ref{EQ_CCRA}) (or (\ref{EQ_LiCCRA})). Following that, two empty sets, $\boldsymbol{\mathcal{R'}}$ and $\boldsymbol{\Omega}$, are established. The former maintains the set of accepted requests, and the latter stores the feasible resource combinations for each request during its iteration. Now, the algorithm iterates through each request in $\boldsymbol{\mathcal{R}}$, starting with the one with the most stringent delay requirement, and keeps track of the feasible allocations of VNF, priority, as well as request and reply paths based on the constraints of (\ref{EQ_CCRA}) (or (\ref{EQ_LiCCRA})). The final steps of each iteration are to choose the allocation with the lowest cost and fix it for the request, as well as to update remaining resources and the set of pending and accepted requests. When there is no pending request, the algorithm terminates.

The complexity of the WF-CCRA algorithm is $O(\mathcal{RVK}|\boldsymbol{\mathcal{P}}|^2)$. Although this approach is significantly more efficient than the B\&B algorithm in terms of complexity (it can be executed within milliseconds), its complexity can be further reduced by restricting the number of valid paths between each pair of nodes to the $\mathcal{P}$ paths with the lowest costs or smallest number of links.
 
\section{Simulation Results}\label{S_SIMULATIONRESULTS}

In this section, the accuracy of the B\&B-CCRA and WF-CCRA methods is numerically investigated. The simulation parameters are listed in Table \ref{tab1}. As long as the problem remains feasible, the values for the remaining parameters can be chosen arbitrarily. Note that the results were obtained on a computer with 8 processing cores, 16 GB of memory, and a 64-bit operating system.

\begin{table}[t!]
\caption{Simulation Parameters}
\begin{center}
\begin{tabular}{|c|c|}
\hline
\textbf{Parameter} & \textbf{Value} \\
\hline
\multirow{2}{*}{number of links ($\mathcal{L}$)} & $\sim \mathcal{U}\{3\mathcal{V}, 5\mathcal{V}\}$, where the \\ & resulted graph is connected. \\
Number of priority levels ($\mathcal{K}$) & $4$ \\
Number of services ($\mathcal{S}$) & $3$ \\
Number of tiers & $3$ \\
Capacity of each link ($\widehat{B_l}$) & $\sim \mathcal{U}\{100, 150\}$ Gbps \\
Cost of each link ($\Xi_l$) & $\sim \mathcal{U}\{10, 20\}$ \\
Capacity of each node ($\widehat{\zeta_v}$) & $\sim 50 \; \mathcal{U}(\alpha, \alpha+1)$ Gbps \\
Cost of each node ($\Psi_v$) & $\sim 50^{\; \mathcal{U}(\alpha, \alpha+1)}$ \\
Priority bandwidth upper bound ($\widehat{f_{l,k}}$) & $\widehat{B_l}/\mathcal{K}$ \\
\hline
\multicolumn{2}{l}{$\alpha$ is the number of tiers minus the tier number of the node.}
\end{tabular}
\label{tab1}
\end{center}
\vspace{-0.7cm}
\end{table}

\begin{figure*}[t!]\centering
\vspace{0.1cm}
\includegraphics[width=6.9in]{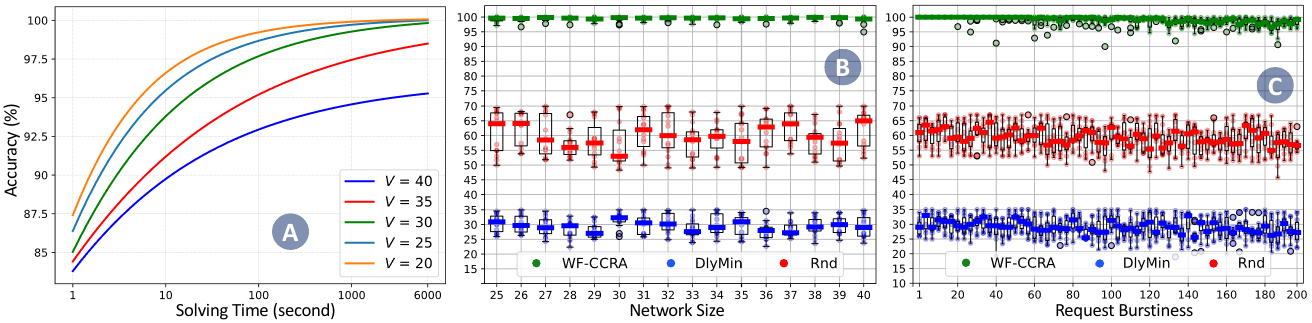}
  \caption{B\&B-CCRA accuracy vs. solving time (A), the accuracy of WF-CCRA, DlyMin, and Rnd vs. network size (B) and request burstiness (C)}
  \label{FIG_SimulationResults}
    \vspace{-0.4cm}
\end{figure*}

The results are illustrated in Fig \ref{FIG_SimulationResults}. The proposed methods are evaluated based on the accuracy of the solutions they provide. Note that the accuracy of a solution for a scenario ($\eta$) is defined as $1-((\eta-\eta^\star)/\eta^\star)$, where $\eta^\star$ is the scenario's optimal solution, which is obtained by solving it with CPLEX 12.10. In Fig \ref{FIG_SimulationResults}-A, the accuracy of B\&B-CCRA is plotted vs. the solving time for five scenarios with different network sizes. In this simulation, the number of requests is set to $200$. As illustrated, the accuracy of B\&B-CCRA starts at 80\% after the first iteration, which is obtained by solving the LP transformation of LiCCRA with CPLEX 12.10 in just a few milliseconds, and increases as the solving time passes, reaching 92\% for all samples after 100 seconds. It proves that this method can be easily applied to provide baseline solutions for small and medium size use cases. However, the accuracy growth is slowed by increasing the network size, which is expected given the problem's NP-hardness and complexity.

In the two remaining sub-figures, the accuracy of WF-CCRA is depicted against the number of requests and network size. In addition, these sub-figures illustrate the outcomes of two more approaches, called DlyMin and Rnd. In the DlyMin method, allocations are performed to minimize delay regardless of other constraints, while Rnd is used to allocate resources randomly to requests. Note that the number of requests in Fig \ref{FIG_SimulationResults}-B is $200$, and the number of network nodes in Fig \ref{FIG_SimulationResults}-C is $20$. For each number of nodes or requests, 50 random systems are formed, and the problem is solved for them using the aforementioned techniques. It is evident that regardless of network size, WF-CCRA has an average accuracy of greater than 99\%, implying that it can be used to allocate resources in a near-optimal manner even for large networks. For different numbers of requests, the average accuracy remains significantly high and greater than 96\%. It does, however, slightly decrease as the number of requests increases, which is the cost of decomplexifying the problem by allocating the resources through separating requests. For the Rnd method, because it consumes the resources of all tiers uniformly, its accuracy is slightly above $50\%$. DlyMin is the least efficient method according to the results. The reason is that this method always utilizes the costly tier-one nodes to minimize E2E delay. In conclusion, it is shown that the WF-CCRA algorithm is capable of efficiently allocating resources for large numbers of requests compared to other approaches.

\section{Conclusion}\label{S_CONCLUSION}

In this paper, the joint problem of communication and computing resource allocation including VNF placement and assignment, traffic prioritization, and path selection considering capacity and delay constraints was studied. The primary goal was to minimize the overall cost. We first formulated the problem as a MINLP model and used a method, named B\&B-CCRA, to solve it optimally. Then, a WF-based approach was developed to find near-optimal solutions in a timely manner. Numerical results demonstrated the efficiency of the proposed methods for large numbers of requests and nodes. As a potential future work, we plan to solve the problem considering the ever-changing characteristics of end-users and infrastructure resources. We intend to devise an online machine-learning approach for real-time adaptation of the allocation strategy to keep the overall cost minimized in such dynamic scenarios. Additionally, we are developing an access control strategy to reduce overall cost over time by predicting future requests.

\section*{Acknowledgment}
This research work is partially supported by the Academy of Finland 6G Flagship, by the European Union’s Horizon 2020 ICT Cloud Computing program under the ACCORDION project with grant agreement No. 871793, and by the European Union’s Horizon 2020 research and innovation program under the CHARITY project with grant agreement No. 101016509. It is also partially funded by the Academy of Finland Project 6Genesis under grant agreement No. 318927.

\bibliographystyle{IEEEtran}
\bibliography{main}

\end{document}